\documentclass[journal=jctcce,manuscript=article,layout=twocolumn]{achemso}
\pdfoutput=1

\usepackage{amsmath} 
\usepackage{hyperref}
\usepackage{graphicx} 
\usepackage{float}
\usepackage{multirow}
\usepackage{amssymb}
\usepackage{soul}
\usepackage[usenames, dvipsnames]{xcolor}
\usepackage[english]{babel}
\usepackage[font={sf,footnotesize},labelfont={bf}]{caption}

\usepackage[version=4]{mhchem}

\renewcommand{\vec}{\bf}

\title{Hyperparameter Optimization for Atomic Cluster Expansion Potentials}

\author{Daniel F. Thomas du Toit}
\author{Yuxing Zhou}
\author{Volker L. Deringer}
\email{volker.deringer@chem.ox.ac.uk}
\affiliation{Inorganic Chemistry Laboratory, Department of Chemistry, University of Oxford, Oxford OX1 3QR, UK}

\begin{document}

\begin{abstract}
Machine-learning-based interatomic potentials enable accurate materials simulations on extended time- and lengthscales. ML potentials based on the Atomic Cluster Expansion (ACE) framework have recently shown promising performance for this purpose. Here, we describe a largely automated computational approach to optimizing hyperparameters for ACE potential models. We extend our openly available Python package, XPOT, to include an interface for ACE fitting, and discuss the optimization of the functional form and complexity of these models based on systematic sweeps across relevant hyperparameters. We showcase the usefulness of the approach for two example systems: the covalent network of silicon and the phase-change material \ce{Sb2Te3}. More generally, our work emphasizes the importance of hyperparameter selection in the development of advanced ML potential models.
\end{abstract}

\maketitle

\section*{Introduction}

Machine learning (ML) approaches to interatomic potential fitting are now widely used in computational chemistry. \cite{Behler2017, Deringer2019, Noe2020, Unke2021, Friederich2021} ML potentials provide surrogate models for quantum-mechanical (QM) potential-energy surfaces at a fraction of the cost, providing longer-timescale and larger-lengthscale simulations than would be accessible with direct QM methods, while maintaining comparable accuracy. As a result, they are now becoming widespread tools to simulate materials and molecules, with applications ranging from the structure of disordered solids \cite{Qamar-23-6, Zhou-23-10} to modeling nuanced effects such as anharmonic phonons \cite{Ouyang-22-3} or non-collinear magnetism in iron, \cite{Rinaldi-24-1} as well as materials not yet synthesized.\cite{Li2022}

Many methods for fitting ML potentials have now been developed, from the early Behler--Parrinello neural-network \cite{Behler2007, Behler-21-03} and Gaussian Approximation Potential (GAP) \cite{Bartok2010, Deringer2021b, Klawohn2023} models to more recent atomic cluster expansion (ACE) potentials, \cite{Drautz2019, Lysogorskiy-21-6, Bochkarev-22-1}, as well as graph-neural-network architectures. \cite{Batzner-22-5, Batatia-22-5, Chen-22-11, Deng-23-09} Each of these methods has different characteristics: for example, graph-based architectures currently define the state-of-the-art in terms of accuracy, but inference (prediction) is still relatively expensive for large systems. \cite{Stocker-22-11, Gardner-24-3}
Our focus herein is on ML potentials that are based on ACE descriptors and fitted using the \texttt{PACEmaker} software by Drautz and coworkers, \cite{Lysogorskiy-21-6, Bochkarev-22-1} which have shown to provide efficient and accurate predictions. \cite{Zuo2020, Lysogorskiy-21-6}

Despite their popularity, ML potentials pose challenges because they largely lack physically motivated functional forms. First, they require high-quality reference data: \cite{BenMahmoud-24-6} recent hand-crafted ML potentials for carbon \cite{Qamar-23-6} and Ge--Sb--Te \cite{Zhou-23-10} each used more than 300,000 atomic environments for training; for potentials encompassing many different elements, the dataset sizes can easily range in the millions. \cite{Batatia-24-3, Yang-24-5} Second, ML potentials require well-chosen hyperparameters---those parameters which must be chosen before fitting starts, and thus determine the nature of the fitting. For mathematically complex ML models in particular, the effects of changing hyperparameters are not always immediately clear. For example, in the ACE approach we use here, the coefficients and exponents of energy contribution terms can be difficult to tune manually. It would therefore be desirable to explore the space of hyperparameters automatically and to optimize them with minimal user input. 

We previously showed how hyperparameter optimization for spectral neighbor analysis potential (SNAP; Ref.~\citenum{Thompson2015}) models can improve their prediction accuracy without increased cost at runtime. \cite{ThomasduToit-23-6} Conversely, with poorly chosen hyperparameters, accuracy and robustness may suffer. 
We note that more widely, key advances have been made in recent years in terms of choosing hyperparameters for ML potentials. \cite{Natarajan2021, Dalbey-21-11, Poelking2022, Park-23-6, ThomasduToit-23-6}

\begin{figure*}
    \centering
    \includegraphics[width=17cm]{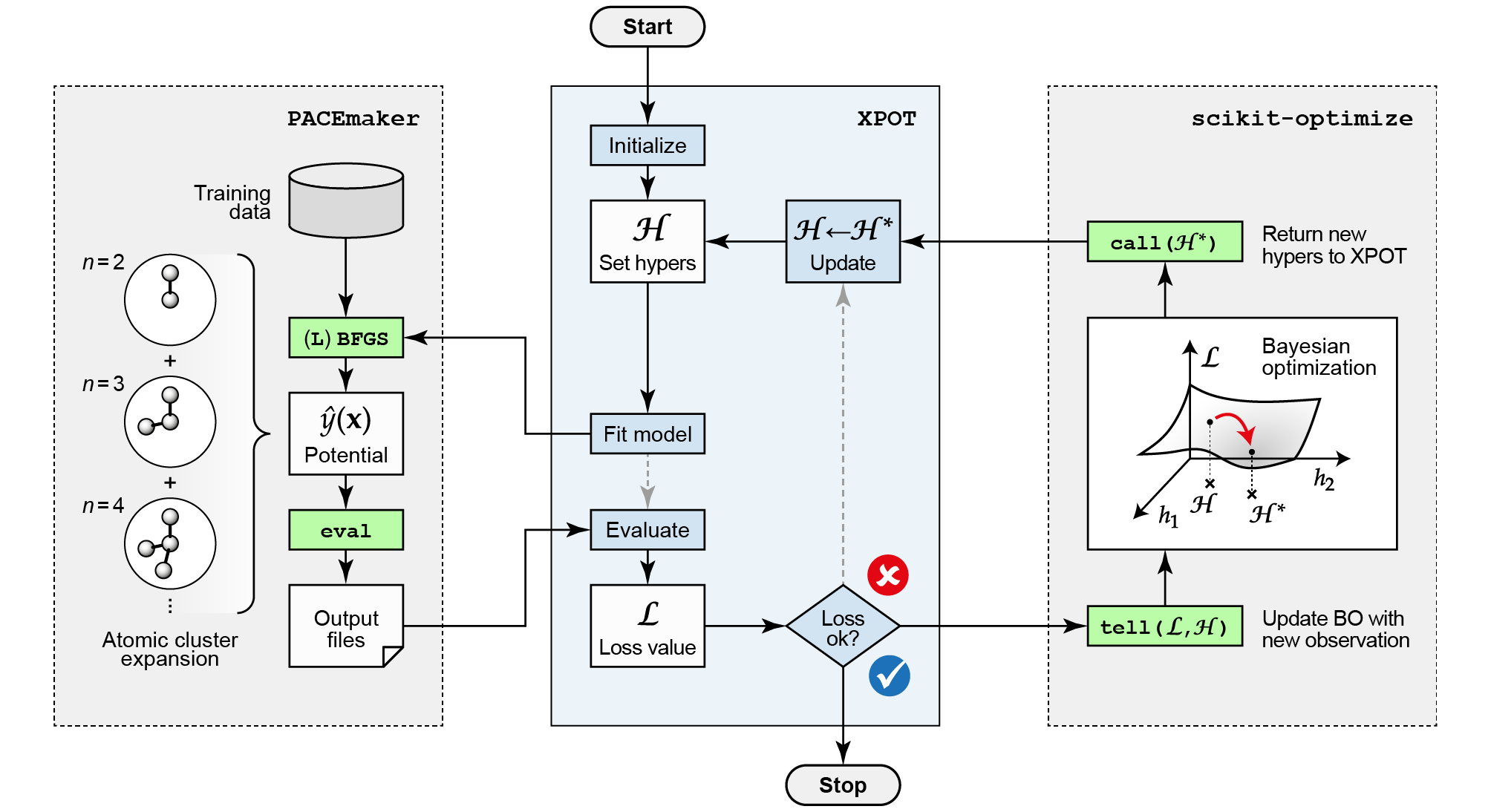}
    \caption{Overview of the methodology for automated optimization of ACE potentials, showing a simplified flowchart of the computational tasks involved. The core functionality of {\tt XPOT} is highlighted in the central box in blue, and external calls to {\tt PACEmaker} ({\em left}) and {\tt scikit-optimize} ({\em right}) are indicated.}
    \label{fig:overview}
\end{figure*}

In this work, we study the effect of hyperparameter optimization for ACE ML potentials, fitted to existing high-quality datasets for silicon \cite{Bartok2018} and \ce{Sb2Te3}. \cite{Zhou-23-10} We investigate both numerical and structural predictions and benchmark the performance of the resulting ACE models. For this purpose, we extended our automated hyperparameter optimization package, ``Cross-platform optimizer for machine-learning interatomic potentials'' (XPOT), \cite{ThomasduToit-23-6} to include support for ACE potentials fitted with \texttt{PACEmaker} (Figure 1). In doing so, we emphasize that XPOT enables efficient transfer between fitting frameworks, here from GAP to ACE---giving access to potentials which are cheaper to run while maintaining comparable accuracy for the same training data.

\section*{Methods}
\label{sec:methods}

\subsection*{Atomic Cluster Expansion}

The atomic cluster expansion (ACE) produces a complete set of basis functions which span the space of local atomic environments, as described in Refs.~\citenum{Drautz2019, Lysogorskiy-21-6, Bochkarev-22-1}. In recent years, the ACE framework has shifted the Pareto front for cost and accuracy of ML potentials \cite{Lysogorskiy-21-6} and has underpinned the development of state-of-the-art interatomic potential models. \cite{Qamar-23-6, Liang2023, Erhard-23-9, Ibrahim2023} For example, a general-purpose ACE potential for carbon \cite{Qamar-23-6} outperforms previously reported models \cite{Deringer2017} in terms of both accuracy and speed, and an ACE model for Si--O was shown to accurately describe complex nanoscale structural features in this key binary system. \cite{Erhard-23-9}

ACE-based ML models are built from general atomic ``properties'', $\varphi_{i}$, expanded across body-ordered functions from the set of neighbors of each atom. Any given property of atom $i$ (index $p$) can therefore be expanded as a function of this atom's local environment:
\begin{equation}
    \label{ace-property}
    \varphi_{i}^{(p)} = \sum_{\textbf{v}} c_{\textbf{v}}^{(p)} \textbf{B}_{i\textbf{v}}
\end{equation}
where the expansion coefficients $c_{\textbf{v}}^{(p)}$ and basis functions ($\textbf{B}_{i\textbf{v}}$) share multi-indices $v$ which describe the list of basis functions in a cluster. The potential energy of the $i$-th atom---typically the prediction target for ML potential models---can be described as a combination of several atomic properties. In the simplest case, the atomic energy, $\varepsilon_{i}$, depends on only a single value:
\begin{equation}
    \label{eq:ace-e-linear}
    \varepsilon_{i} = \varphi_{i}^{(1)},
\end{equation}
which leads to a linear model. However, ACE potentials are not limited to using a single property: introducing nonlinear behavior through a Finnis--Sinclair-like embedding, for example, results in
\begin{equation}
    \label{eq:ace-e-fs}
    \varepsilon_{i} = \varphi_{i}^{(1)} + \sqrt{\varphi_{i}^{(2)}}
\end{equation}
where the two properties are generalizations of the pairwise repulsion and density of the widely used Finnis--Sinclair potential. \cite{Finnis-1984} This approach can be further extended to an arbitrary number of $P$ atomic properties ($p = 1, \ldots, P$) which all enter as arguments into a nonlinear function, denoted $F$:
\begin{equation}
    \label{eq:ace-e-gen}
    \varepsilon_{i} = F( \varphi_{i}^{(1)}, \ldots , \varphi_{i}^{(P)})
\end{equation}
where $i$ again is the atomic index. For example, in Ref.~\citenum{Erhard-23-9}, it was shown that a complex functional form ($P=8$) outperformed linear and Finnis--Sinclair type embeddings for ACE models for the Si--O system while only leading to a $\approx 15\%$ increase in computational requirements. A full systematic study of the number, exponents, etc.\ of the terms that enter $F$ is yet to be reported, and it is likely that there is some scope for optimizing nonlinear ACE models in this regard.

\subsection*{XPOT}
\label{sec:xpot}

We use our Python package, XPOT, \cite{ThomasduToit-23-6} to fit optimized ACE models. XPOT automates the optimization of hyperparameters within ranges specified by the user, by iteratively minimizing a combined energy and force loss function which is defined as
\begin{equation}
 \label{eq:loss}
 \mathcal{L} = \frac{\alpha}{[E]} \times \mathcal{L}_{\rm{E}} +  \frac{(1-\alpha)}{[\vec{F}]} \times \mathcal{L}_{\rm{F}}.
\end{equation}
In this expression, square brackets indicate the units (in the present work, we use energies in eV $n_{\rm at}^{-\frac{1}{2}}$ and forces in eV~\AA$^{-1}$~respectively), emphasizing that the loss function, $\mathcal{L}$, itself is dimensionless.
By adjusting the relative energy and force loss weighting, controlled by the single parameter $\alpha$, the potential fit can be optimized toward the desired characteristics. 

The energy contribution to the loss function, $\mathcal{L}_{\rm{E}}$, is given by
\begin{equation}
 \label{eq:loss_energy}
 \mathcal{L}_{\rm{E}} = \sqrt{\frac{1}{N_{\rm cells}}\sum_{i \in {\textbf D}_{\rm{val}}} 
 \left(\frac{\hat{E}_i-E_i}{\sqrt{n_{{\rm at},i}}}\right)^2}
\end{equation}
where $N_{\rm{cells}}$ is the number of structures in the validation set, ${\textbf D}_{\rm{val}}$, and $n_{\mathrm{at},i}$ is the number of atoms in the $i$-th structure from that set. The energy prediction of the ACE model is $\hat{E}_{i}$, whereas the ``correct'' reference energy in the validation set is $E_{i}$. The difference between the two is divided by $\sqrt{n_{\mathrm{at},i}}$---not $n_{\mathrm{at},i}$---in order to weight the contribution of energy errors from each structure equally, regardless of the number of atoms in any given validation-set structure. \cite{Morrow-23-03}

The force contribution to the loss function,
\begin{equation}
 \label{eq:loss_force}
 \mathcal{L}_{\rm{F}} = \sqrt{\frac{1}{N_{\rm at}}\sum_{j \in \textbf{D}_{\rm{val}}}\left|\hat{\vec F}_{j} - {\vec F}_{j} \right|^2},
\end{equation}
is determined by measuring the magnitude of the error between the predicted and reference force vectors. Since the forces are per-atom quantities, the overall error is averaged over the total number of atoms in the dataset, $N_{\rm{at}}$.

\subsection*{Implementation of ACE Support in XPOT}

Figure \ref{fig:overview} lays out the workflow of XPOT, outlining the distribution and ordering of tasks within the hyperparameter optimization loop. This figure specifically relates to {\tt PACEmaker} and ACE potentials, but the XPOT workflow is similar for other fitting methods. XPOT interfaces both to potential fitting software (left-hand side in Figure \ref{fig:overview}) and to the Bayesian optimization (BO) interface of \texttt{scikit-optimize} (right-hand side). The latter supports discrete and continuous variable optimization, as discussed in Ref.~\citenum{ThomasduToit-23-6}. The user-defined hyperparameter ranges are parsed, before potentials are fitted iteratively, and new hyperparameters are chosen through BO predictions until the loss reaches the desired value, or the maximum number of iterations is reached. A full description of the XPOT code can be found in Ref.~\citenum{ThomasduToit-23-6}.

To begin the process, the user initializes the optimization in Python, specifying an input script containing the hyperparameters to be optimized. XPOT parses this input and uses \texttt{scikit-optimize} to set values for each iteration. Then, \texttt{PACEmaker} is invoked, fitting the ACE potential model. XPOT parses the output files and converts the prediction errors into a single loss function, $\mathcal{L}$ (see Eq.~\ref{eq:loss}). At this stage, if optimization is determined to be finished, XPOT exits, otherwise calling another estimation of the loss surface and updating the hyperparameters to continue the  process.

While the optimization process in XPOT is agnostic to the specific fitting methodology, file parsing between XPOT and the desired fitting software must be implemented. 
Here, to reduce compute-time and installation requirements, evaluation of the loss function is performed using the detailed per-structure predictions provided by \texttt{PACEmaker}. As a result, unlike XPOT's interface to \texttt{fitsnap}, \texttt{LAMMPS} is not required for optimization of ACE potentials.
GPU fitting with \texttt{PACEmaker} is supported in XPOT, and can be controlled by specifying the desired CUDA device at runtime using environment variables.

\subsection*{Datasets}
\label{sec:datasets}

We briefly describe the datasets that we use in the present study for fitting and validating potentials. For details, we refer to the cited original publications; for an overview of validation techniques for ML potentials more generally, we refer to Ref.~\citenum{Morrow-23-03}.

In the case of silicon, we use three different datasets for specific purposes. First, we take the training dataset for the Si-GAP-18 general-purpose potential by Bartók et al. \cite{Bartok2018} as an example of a highly-developed and largely handcrafted dataset. We fit ACE models to this dataset using XPOT, and we use the corresponding test set for validation (both taken from Ref.~\citenum{Bartok2018}). Second, we use DFT-labeled snapshots from a GAP-MD simulation described in Ref.~\citenum{George2020} (referred to as ``MQ-MD'' in the following). This dataset contains diamond-type supercells with vacancy defects, liquid, and amorphous structures, as well as transitions between these phases. The structures were generated with Si-GAP-18-driven MD and subsequently labeled with single-point DFT computations. \cite{George2020} Finally, we test our XPOT-fitted models on random structure search (RSS) configurations from Ref.~\citenum{Morrow-22-9}. These allow us to test the potentials' ability to describe structures different from those on which they were trained.

\begin{figure}[h]
   \centering
   \includegraphics[width=\linewidth]{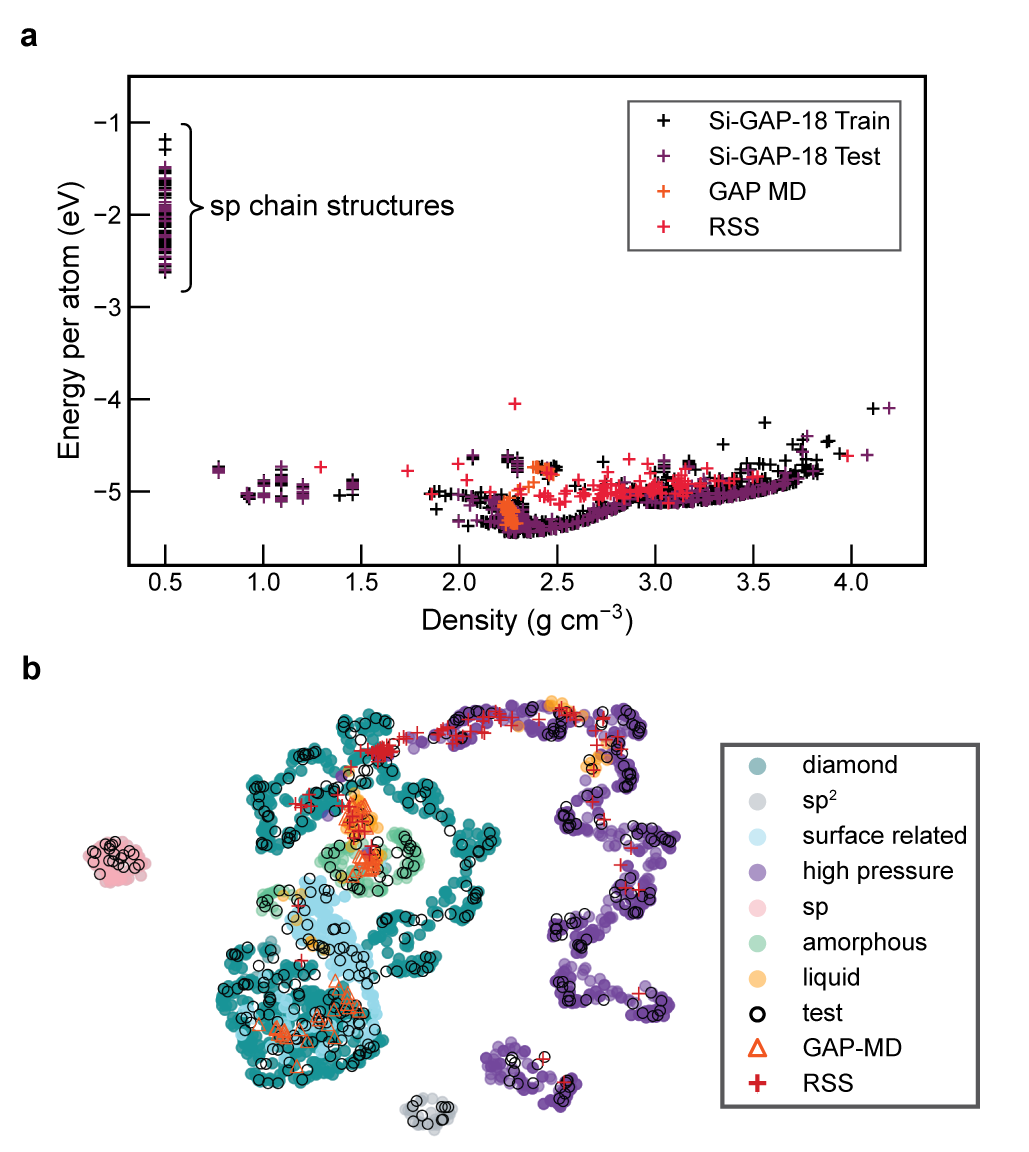}
   \caption{Visualization of the data for Si used in this work. (\textbf{a}) An energy--density plot showing the heterogeneous nature of the Si-GAP-18 training dataset (Ref.~\citenum{Bartok2018}), with the ``sp'' chain structures highlighted. (\textbf{b}) UMAP embedding of the averaged ACE vectors for structures in the silicon datasets, highlighting the overlap in local atomic environments between structures. The categories are partly simplified compared to Ref.~\citenum{Bartok2018}: here, for example, ``diamond'' also includes defective diamond-type Si structures.}
   \label{fig:struct-maps}
\end{figure}

In Figure \ref{fig:struct-maps}a, we characterize the various training and testing sets used by plotting the energy against the mass density for each structure. The bulk structures mainly have densities relevant to amorphous, crystalline, and liquid silicon, with sp-like chain structures constituting the dataset entries at very low density and high energy.
Despite the difference in the composition of the RSS, MQ-MD, and Si-GAP-18 test datasets, the structures seem broadly related when compared in this plot.

In Figure \ref{fig:struct-maps}b, we show a similarity map, obtained by UMAP dimensionality reduction \cite{UMAP} on the ACE vectors of the atomic environments averaged for each structure. This map indicates distinct regions, mirroring the varied structure types included in the dataset. 
By color-coding the points in the map according to the configuration types defined in the Si-GAP-18 training dataset, \cite{Bartok2018} we can visualize the isolated nature of the low-coordinate (``sp'' and ``sp$^2$'') and high-pressure structures ($\beta$-Sn-type and simple hexagonal), and the relation between the various phases included. Additionally, we can visualize the coverage of the various test sets and their similarity to the training data. The RSS dataset includes entries that (although clearly higher in energy; cf.\ Figure \ref{fig:struct-maps}a) resemble diamond-like, high-pressure, and liquid-like structures. Finally, we note that all surface-related structures (slab models) in the Si-GAP-18 training set are based on the diamond-type form, and so the similarity of these structures to bulk diamond-type Si is expected.

The dataset used in our study of Sb$_2$Te$_3$ is taken from the GST-GAP-22 dataset. \cite{Zhou-23-10} In this case, we remove all structures which include Ge, yielding a dataset for the binary Sb--Te system. 
We randomly split this dataset into train and test data (80:20), for each configuration type as defined in the dataset, such as ``{\tt crystalline}", ``{\tt aimd}", and ``{\tt liquid}". This way, we can ensure that the validation data span a wide range of structure types, and reduce the possibility of optimizing overly towards a single subgroup of the dataset.

\section*{Results and Discussion}\label{sec:results}

\begin{table*}[tp!]
 \caption{\label{tab:si_test} Energy and force RMSE values of silicon potentials, evaluated on three different test sets. For the XPOT-ACE models, the number of atomic properties, $P$, as defined in Eq. \ref{eq:ace-e-gen}, ranges from 1 to 4. An ``F'' denotes a potential where the number of functions was optimized by XPOT. XPOT-ACE-6827 is an optimized model using the same number of radial basis functions as the linear ACE potential fitted by Lysogorskiy et al. \cite{Lysogorskiy-21-6} (denoted as ``REF-ACE'' here). The MD speed is given relative to that of Si-GAP-18.}
 \fontsize{9.5pt}{10.25pt}\selectfont
 \begin{tabular}{lcc|ccc|ccc|c}
 \hline
 \hline
  & & & \multicolumn{3}{c}{Energy RMSE} & \multicolumn{3}{c}{Force RMSE} \\
  & & & \multicolumn{3}{c}{(meV at.$^{-1}$)} & \multicolumn{3}{c}{(meV \AA$^{-1}$)} \\
  \cline{2-5} \cline{6-9}
  & $P$ & \# Func. & Si-GAP-18-Test\cite{Bartok2018} & MQ-MD\cite{George2020} & RSS\cite{Morrow-22-9} & Si-GAP-18-Test & MQ-MD & RSS & MD speed\\

 \hline
 XPOT-ACE-1 & 1 & 3,000 & 3.5 & 5.1 & 27.1 & 69 & 105 & 158 & 47 \\
 XPOT-ACE-2 & 2 & 3,000 & 2.5 & 5.0 & 23.1 & 63 & 97 & 150 & 46 \\
 XPOT-ACE-3 & 3 & 3,000 & 318 & 5.2 & $> 10^6$ & 300 & 98 & $> 10^8$ & 45 \\
 XPOT-ACE-4 & 4 & 3,000 & 4.8 & 5.5 & 62.6 & 63 & 99 & 274 & 43 \\
 \hline
 XPOT-ACE-3F & 3 & 2,000 & 4.6 & 4.1 & 20.5 & 65 & 97 & 139 & 65 \\
 XPOT-ACE-4F & 4 & 1,625 & 2.5 & 5.4 & 72.5 & 64 & 100 & 187 & 80 \\
  \hline
 XPOT-ACE-6827 & 1 & 6,827 & 3.0 & 4.4 & 34.5 & 63 & 104 & 179 & 16 \\
 REF-ACE~\cite{Lysogorskiy-21-6} & 1 & 6,827 & 3.2 & 4.3 & 42.1 & 77 & 124 & 175 & 16 \\
 Si-GAP-18~\cite{Bartok2018} & --- & --- & 1.6 & 8.5 & 34.9 & 83 & 139 & 177 & 1 \\
 \hline 
 \hline
 \end{tabular}
\end{table*}

\subsection*{Silicon (I): Numerical Performance}

\begin{figure}[h]
   \centering
   \includegraphics[width=\linewidth]{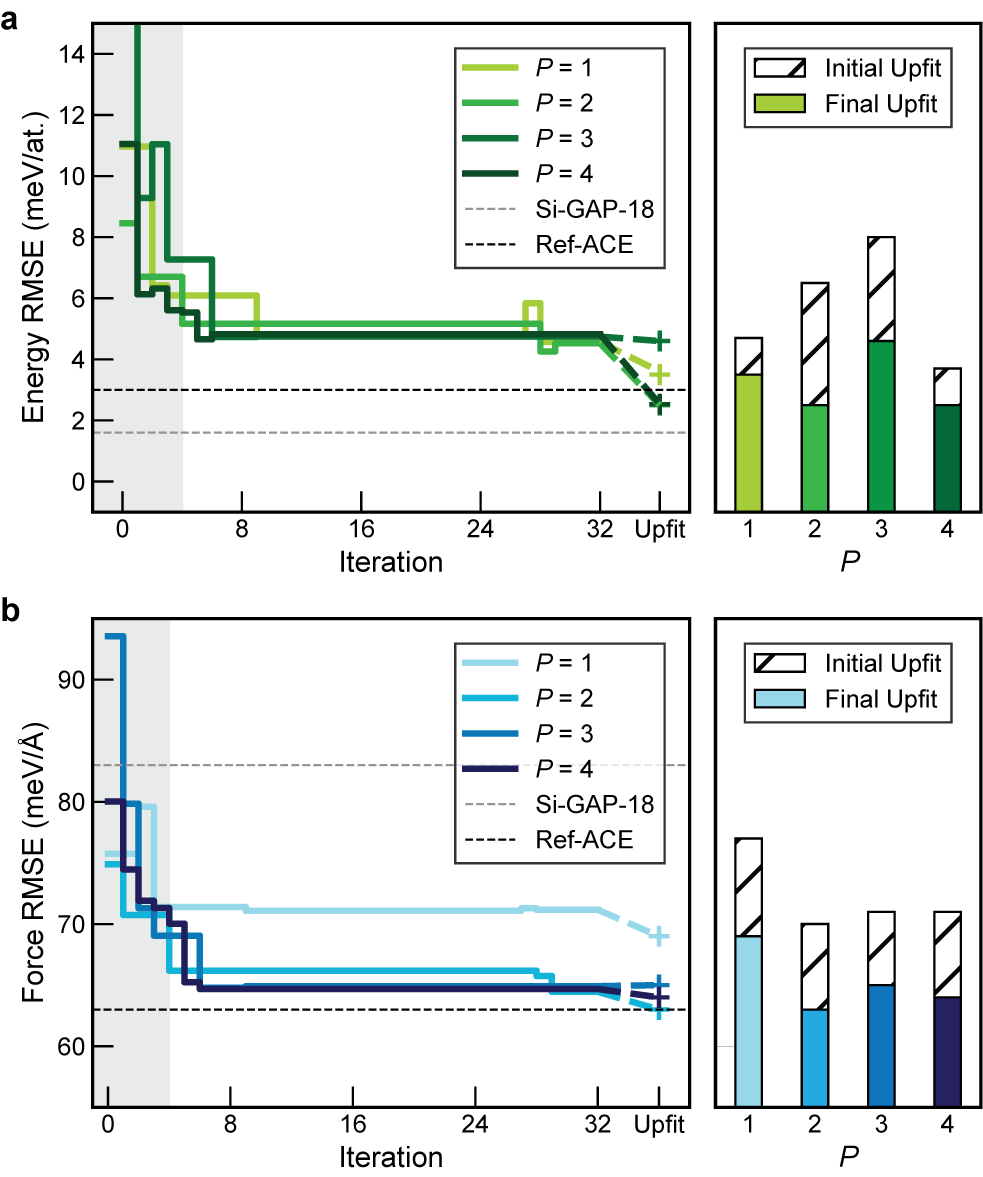}
   \caption{Evolution of energy and force errors for silicon ML potentials through iterative optimization using XPOT. We show (\textbf{a}) the per-atom energy RMSE and (\textbf{b}) the force component RMSE across iterations. Errors are evaluated on the Si-GAP-18 test set from Ref.~\citenum{Bartok2018}. On the left, the gray region indicates the initial sampling stage of optimization (see Ref.~\citenum{ThomasduToit-23-6}). The bar charts show the errors on the test set for ``upfitted'' potentials, as described in the text: both for the best potentials from the initialization protocol (hatched), and the best potentials after Bayesian optimization (solid). The data for $P=3$ and $P=4$ refer to the `-3F' and `-4F' potentials from Table \ref{tab:si_test}, respectively.}
   \label{fig:si_e_f_evolution}
\end{figure}

We fitted ACE potentials to the general-purpose Si-GAP-18 training dataset from Ref.~\citenum{Bartok2018}. Previously, Lysogorskiy et al.\ fitted a linear potential to the same data, \cite{Lysogorskiy-21-6} and several non-linear models ($P > 1$) have been fitted recently for other materials. \cite{Qamar-23-6, Erhard-23-9} Here, we create ACE models over the range of $1 \le P \le 4$, using fewer basis functions compared to Ref.~\citenum{Lysogorskiy-21-6}, in order to improve efficiency at runtime. Our XPOT-ACE-2 model achieves accuracy comparable to the potential fitted in Ref.~\citenum{Lysogorskiy-21-6} (``Ref-ACE'', Figure \ref{fig:si_e_f_evolution}), taking advantage of optimized hyperparameters and non-linear ACE models for improved efficiency.

XPOT optimizes hyperparameters by minimizing the combined loss, $\mathcal{L}$ (Eq. \ref{eq:loss}). As such, XPOT is guided by changes in the errors on the specific validation set (here, the Si-GAP-18 test set \cite{Bartok2018}). As the MQ-MD and RSS datasets are {\em not} used as targets in XPOT optimization, we leverage them as distinct benchmarks for accuracy and robustness. The MQ-MD dataset from Ref.~\citenum{George2020} is similar in character to the parts of the Si-GAP-18 test set, allowing further accuracy tests on larger structures. In contrast, the RSS dataset contains randomly generated structures which are typically higher in energy (Figure \ref{fig:struct-maps}a). No explicit RSS structures are present in either the Si-GAP-18 test or train sets. Using these two datasets, we can validate the performance of the Si-GAP-18 test set as a general-purpose optimization target.

For each value of $P$, we optimized up to 6 hyperparameters at once, and performed 32 fitting iterations. The first four of these were initialization fits (shaded areas in Figure \ref{fig:si_e_f_evolution}), where a pseudo-random sampling method is used to determine hyperparameters, before Bayesian Optimization (BO) is applied for the remaining iterations. For all models, we optimized $\varphi_{i}$ exponents, cutoff, radial basis function (including \texttt{radbaseparameter}), and \texttt{dcut} (the smoothing distance at the outer limit of the cutoff). We used universal structure weighting---unlike REF-ACE, which includes weightings based on the type of structure in the training set. \cite{Lysogorskiy-21-6} This was done to show that even without a ``bespoke" approach to weighting dataset entries, hyperparameter optimization could improve ML potentials. In future work, we hope to study the effect of weighting techniques to further improve fitting for smaller datasets.

After the optimization, the best potential for each optimization sweep was ``upfitted''. By this term, we mean the approach of continuing the fitting process from an existing potential. The accuracy of ACE models can potentially be improved in this way: by first using a higher value of $\kappa$ to emphasize forces in the fit, before refining energy accuracy with low $\kappa$ values. \cite{bochkarev-personal} This process is analogous to the \texttt{--swa} protocol used in MACE fitting.\cite{mace-documentation} Herein, we first fitted using $\kappa = 0.8$, before upfitting the same potential with $\kappa = 0.02$. This process was the same for all XPOT-ACE potentials in the present work. After the final potentials were upfitted, we performed numerical validation on two external datasets (see Methods section).

Initially, we fitted potentials with 3,000 functions each for all values of $P$, but found an increased likelihood of overfitting for $P \ge 3$. Table \ref{tab:si_test} includes results from optimizing potentials for $P=3$ and $P=4$ with 3,000 functions, highlighting the reduced accuracy and robustness compared to the $P \le 2$ potentials. 
Specifically, the XPOT-ACE-3 potential has become much worse during the upfitting procedure, producing highly unreasonable energy and force errors on the RSS test set. When analyzing the errors per structure for the upfitted XPOT-ACE-3 potential further, it was found that the increased error on the Si-GAP-18 test set comes from two individual structures. We emphasize that these issues would not have become apparent from testing only on the MQ-MD dataset, and instead a more wide-ranging analysis including RSS test data reveals that the potential is not sufficiently robust. The non-upfitted potential result from XPOT has a testing error of 7.9 meV at.$^{-1}$---note that whilst this seems to be an acceptable result, it is already $4 \times$ the training error, and thus this potential as well appears to be overfitted.

\begin{figure}[ht!]
   \centering
   \includegraphics[width=0.95\linewidth]{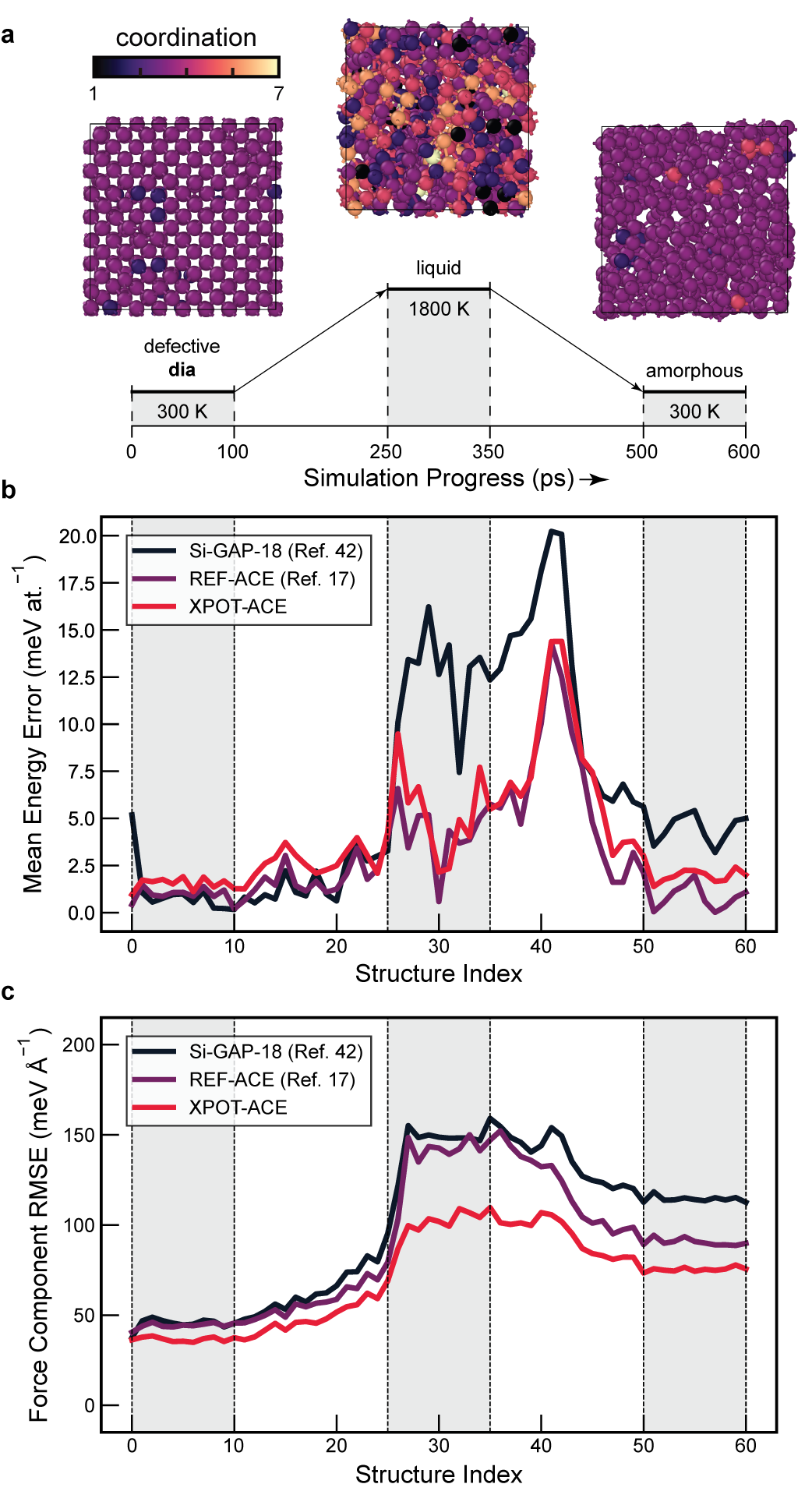}
   \caption{Accuracy of silicon ML potentials evaluated on DFT-labeled snapshots from a Si-GAP-18-driven melt--quench simulation reported in Ref.~\citenum{George2020}, and visualized in the style of that prior work. (\textbf{a}) An overview of the constant-pressure simulation protocol, adapted from Ref.~\citenum{George2020}. The images show the three classes of structure seen at each stage of the simulation, color-coded according to coordination number. DFT snapshots were computed every 10 ps throughout the simulation. (\textbf{b}) Energy errors compared to DFT snapshots along the simulation trajectory. (\textbf{c}) Force errors for the same structures. XPOT-ACE outperforms all other potentials studied here in terms of force errors, but trails in energy errors to the ACE from Ref.~\citenum{Lysogorskiy-21-6}.}
   \label{fig:si_gapmd_benchmark}
\end{figure}

\begin{table}[t]
\caption{\label{tab:si_quench} Energy per-atom differences of quenched a-Si structures relative to (diamond-type) crystalline silicon. Structures were quenched and relaxed using the ML potentials in the top row, comparing the differences between XPOT-ACE and REF-ACE. \cite{Lysogorskiy-21-6} The standard deviation across the sampled structures is reported using parentheses.}
\small
\begin{tabular}{ccccc}
\hline
\hline
Quench rate & \multicolumn{4}{c}{$\Delta E$ (meV at.$^{-1}$)} \\ 

(K $\mathrm{s^{-1}}$) &  \multicolumn{4}{c}{XPOT-ACE Quenched }\\
\hline
 & XPOT & REF & GAP & DFT \\
\cline{2-5}
$10^{14}$                   & $205 (6)$ & $209 (6)$ & $215 (7)$ & $210 (7)$ \\
$10^{13}$                   & $183 (4)$ & $187 (4)$ & $192 (4)$ & $185 (5)$ \\
$10^{12}$                   & $158 (3)$ & $161 (3)$ & $166 (4)$ & $156 (4)$ \\
$10^{11}$                   & $140 (4)$ & $142 (5)$ & $146 (5)$ & $137 (5)$ \\
\hline
 & \multicolumn{4}{c}{REF-ACE Quenched} \\
\hline
 & XPOT & REF & GAP & DFT \\
\cline{2-5}
$10^{14}$                   & $211 (4)$ & $208 (4)$ & $218 (3)$ & $217 (6)$\\
$10^{13}$                   & $190 (4)$ & $187 (3)$ & $197 (4)$ & $185 (5)$ \\
$10^{12}$                   & $160 (4)$ & $158 (3)$ & $166 (4)$ & $155 (5)$\\
$10^{11}$                   & $147 (4)$ & $145 (4)$ & $153 (4)$ & $144 (5)$ \\
\hline
\hline
\end{tabular}
\end{table}

As such, we used XPOT to optimize the number of functions for the $P \ge 3$ potentials as an additional variable hyperparameter, from 500 up to an upper limit of 2,000. By reducing the maximum number of functions, we restricted the flexibility of these models, to optimise towards more robust behaviour. We note that XPOT-ACE-3F uses the maximum number of functions (2,000), while XPOT-ACE-4F is optimized to 1625 functions. These optimized potentials are denoted with an ``F'' suffix and the results are included in Table \ref{tab:si_test}.
In addition to having improved numerical accuracy and efficiency, these potentials were able to complete melt--quench simulations for 4,096-atom cells (Table \ref{tab:si_quench}), which the XPOT-ACE-3 and XPOT-ACE-4 potentials were not. However, as the optimization still occurs for the Si-GAP-18 test set, a good performance on RSS data is still not guaranteed, as shown by the high RSS error for XPOT-ACE-4F.

The evolution of energy and force errors is visualized in Figure \ref{fig:si_e_f_evolution}, including the upfitting procedure which is undertaken at the end of the optimization process, as described above. For comparison, we also upfitted the best potentials from the initialization phase, to assess whether the BO part of the process brings additional benefit. We see that across all values of $P$, the final optimized potentials are more accurate on the Si-GAP-18 test set used to evaluate $\mathcal{L}$.

The upfitting process for the initial potentials does not always improve testing error, despite the training error reducing throughout the fitting process. This suggests that the initial hyperparameters selected are not providing a generalizable description of silicon, indicated by relatively poor predictive accuracy on unseen environments.  Additionally, not all upfitted initial potentials were found to be stable in MD simulations up to 1,800 K. We further quantify the accuracy of these potentials in the Supporting Information.

For the nonlinear XPOT-ACE-4F model, we observe that the robustness and accuracy on the test set are sufficient, but it is prone to large errors on other datasets---that is, the model appears to be overfitted. This finding again demonstrates that, as we discussed in Ref.~\citenum{ThomasduToit-23-6}, the make-up of the validation set is paramount in optimizing a potential towards the desired characteristics. Additionally, setting bounds for the flexibility of the potential reduces the likelihood of overfitting, but the composition of the validation set still directly affects the loss function, and thus the minimum in hyperparameter space towards which the potentials are optimized.

Taking into account the results above, XPOT-ACE-2 seems to offer the best combination of accuracy and robustness among the entries of Table \ref{tab:si_test}. We therefore take this potential forward for further testing, and from here onwards we refer to it simply as ``XPOT-ACE''.

\begin{figure*}[ht]
   \centering
   \includegraphics[width=\linewidth]{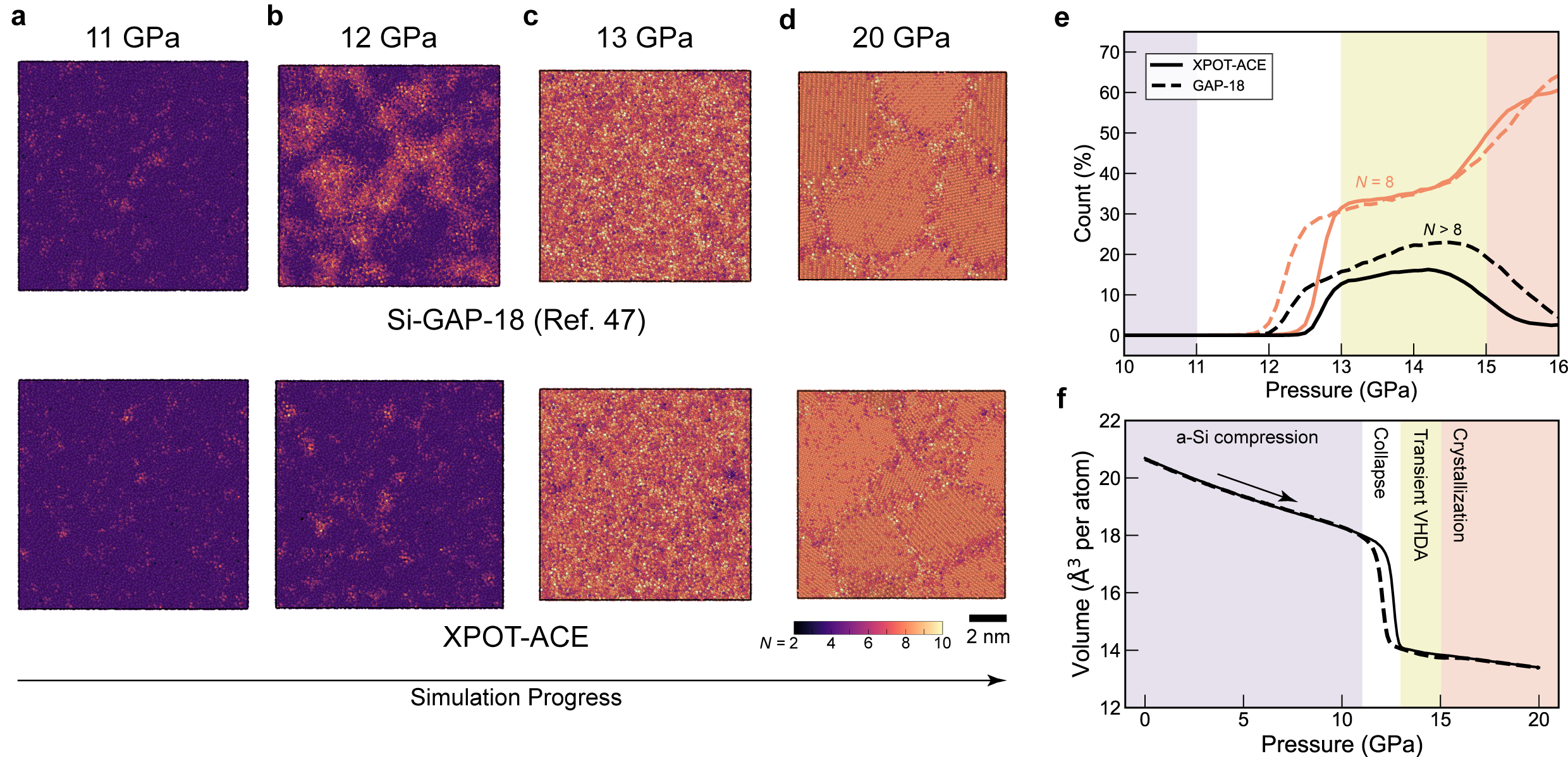}
   \caption{(\textbf{a--d}) Simulations of amorphous silicon under isothermal compression with both XPOT-ACE and Si-GAP-18, \cite{Bartok2018} similar to the simulations reported in Ref.~\citenum{Deringer2021} from which data for the GAP simulation are taken. Both potentials predict a collapse into VHDA occurring between 12--13 GPa from which simple hexagonal crystallites then form. (\textbf{e}) Coordination numbers as a function of pressure in the trajectories (determined by counting neighbors up to 2.85 \AA). The initial increase in $N > 8$ atoms corresponds to the formation of the VHDA phase before crystallization occurs. (\textbf{f}) Volume against pressure during the simulations.}
   \label{fig:si_pressure}
\end{figure*}

\subsection*{Silicon (II): Physics-Guided Validation}

In addition to reporting the overall RMSE for the MQ-MD test set, we plot the errors for each structural snapshot from this dataset in Figure \ref{fig:si_gapmd_benchmark}. Doing so provides a more nuanced view of the types of configurations for which the potential shows higher or lower errors. \cite{George2020} We show that XPOT-ACE improves the force predictions compared to REF-ACE, whilst having very slightly higher errors (to within 1 meV at.$^{-1}$). We presume that due to the uniform structure weighting, our potential is comparatively most improved in the higher-energy liquid state, while improvements for crystalline and amorphous structures are reduced. Notably, the higher weighting of the crystalline phases for existing potentials \cite{Bartok2018, Lysogorskiy-21-6} offers improved accuracy for crystalline configurations, whereas our potential still improves the force accuracy for those.

Across the simulation snapshots, the same structures are resulting in ``spiking'' (high error) energies across all three models. This suggests some underlying characteristic of the Si-GAP-18 training dataset results in these less accurate predictions across models, especially on freezing of the liquid state (which is not strongly represented in the training data). For the latter, all three potentials show increased energy errors. The fluctuations are much reduced in force predictions, but there is still a visible ``bump'' in these predictions for the same structures.

Next, we quenched 500-atom randomized structures at fixed rates from $10^{14}$ to $10^{11}$ K $\mathrm{s^{-1}}$. Both XPOT-ACE and REF-ACE potentials were used and we compared the energies of the structures produced. On top of labeling these structures with XPOT-ACE, REF-ACE, and Si-GAP-18, we compute DFT energies for the structures, to quantify the predictive accuracy of all three potentials (akin to our studies of quenched \ce{SiO2} in Ref.~\citenum{Erhard2022}). For each rate, 5 random structures were quenched, with both mean and standard deviation reported in Table \ref{tab:si_quench}. We tested faster quench rates, but found that rates of $10^{14}$ and $10^{15}$ K $\mathrm{s^{-1}}$ led to structures within 2 meV $\mathrm{at.^{-1}}$ of each other for all models. We therefore do not include results for quench rates of $> 10^{14}$ K $\mathrm{s^{-1}}$ in Table \ref{tab:si_quench}.

Although the energies of structures for $10^{12}$ and $10^{13}$ K $\mathrm{s^{-1}}$ are very similar, XPOT-ACE predicts structures with slightly lower energies for $10^{11}$ K $\mathrm{s^{-1}}$, suggesting a more relaxed and therefore more stable a-Si structure. All potentials provide a good match to DFT. When compared to our findings in Table \ref{tab:si_test} and Figure \ref{fig:si_gapmd_benchmark}, this is in line with our expectations.

Finally, we performed structural validation tests for the compression of silicon, as described in Ref.~\citenum{Deringer2021}. This test was comprised of compressing a 100,000 atom low density amorphous silicon model up to 20 GPa. The pressurization rate is 0.1 GPa ps$^{-1}$, and the temperature is held at 500 K. We observed similar behavior to what was seen in simulations using Si-GAP-18, \cite{Deringer2021} whereby low-density amorphous (LDA) silicon upon compression collapses into a very-high-density amorphous (VHDA) phase at $\approx 12$ GPa, and subsequently simple-hexagonal (sh) crystallites nucleate and grow. This result demonstrates that our XPOT-ACE model has learned much of the same behavior as Si-GAP-18 from the training set.

The formation of the VHDA phase in the compression simulation the using XPOT-ACE potential occurred at a pressure 0.5 GPa higher than that observed for Si-GAP-18, as shown by the relative lack of highly-coordinated silicon atoms in Figure \ref{fig:si_pressure}b---an effect which is corroborated by Figure \ref{fig:si_pressure}e, where the percentage of $N = 8$ atoms does not rise substantially until 12.5 GPa, in contrast to the Si-GAP-18 simulation. Aside from this slightly delayed formation of VHDA, the potentials predicted almost identical densities throughout the simulation, and show behavior that is consistent with experiments for both VHDA formation \cite{McMillan2005} and crystallization. \cite{Pandey-11-6}

\subsection*{Antimony Telluride}

\begin{figure}[h!]
   \centering
   \includegraphics[width=0.95\linewidth]{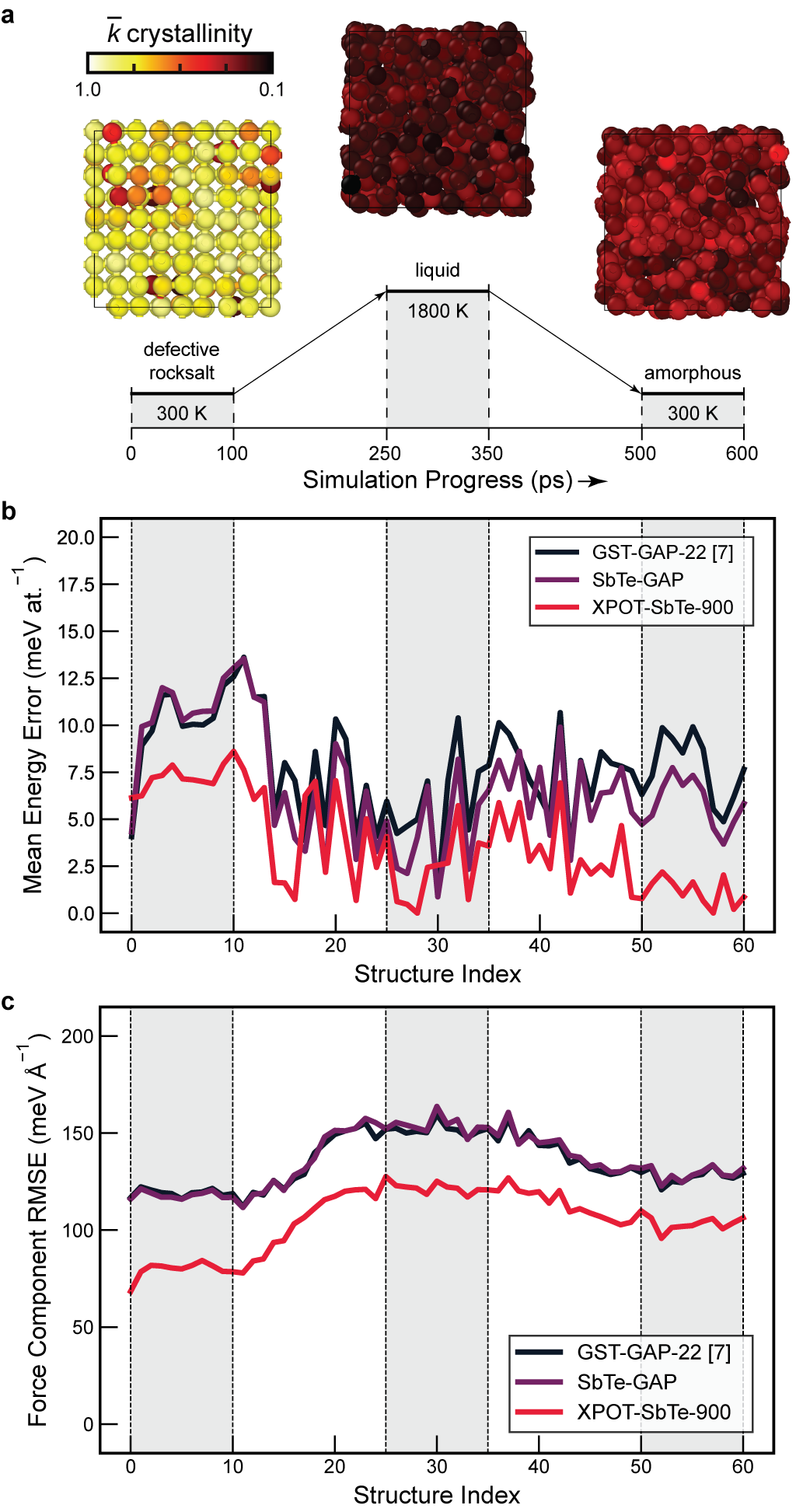}
   \caption{ Prediction accuracy of \ce{Sb2Te3} ML potentials across DFT-labeled snapshots from a GST-GAP-22 melt-quench simulation. (\textbf{a}) An overview of the MD-based benchmark protocol created in a way similar to Ref.~\citenum{George2020}, now for \ce{Sb2Te3}. Structures show the three classes of structures seen at each stage of the simulation, color-coded according to a per-atom crystallinity measure. \cite{Bartok2013, Xu-22-1} Snapshots were labeled with DFT every 10 ps throughout the simulation. (\textbf{b}) Energy errors compared to DFT snapshots across the simulation trajectory. (\textbf{c}) Force errors across the same structures. XPOT-ACE outperforms both GAP potentials on both energy and force errors.}
   \label{fig:sbte_gapmd}
\end{figure}

\begin{figure*}[]
   \centering
   \includegraphics[width=\linewidth]{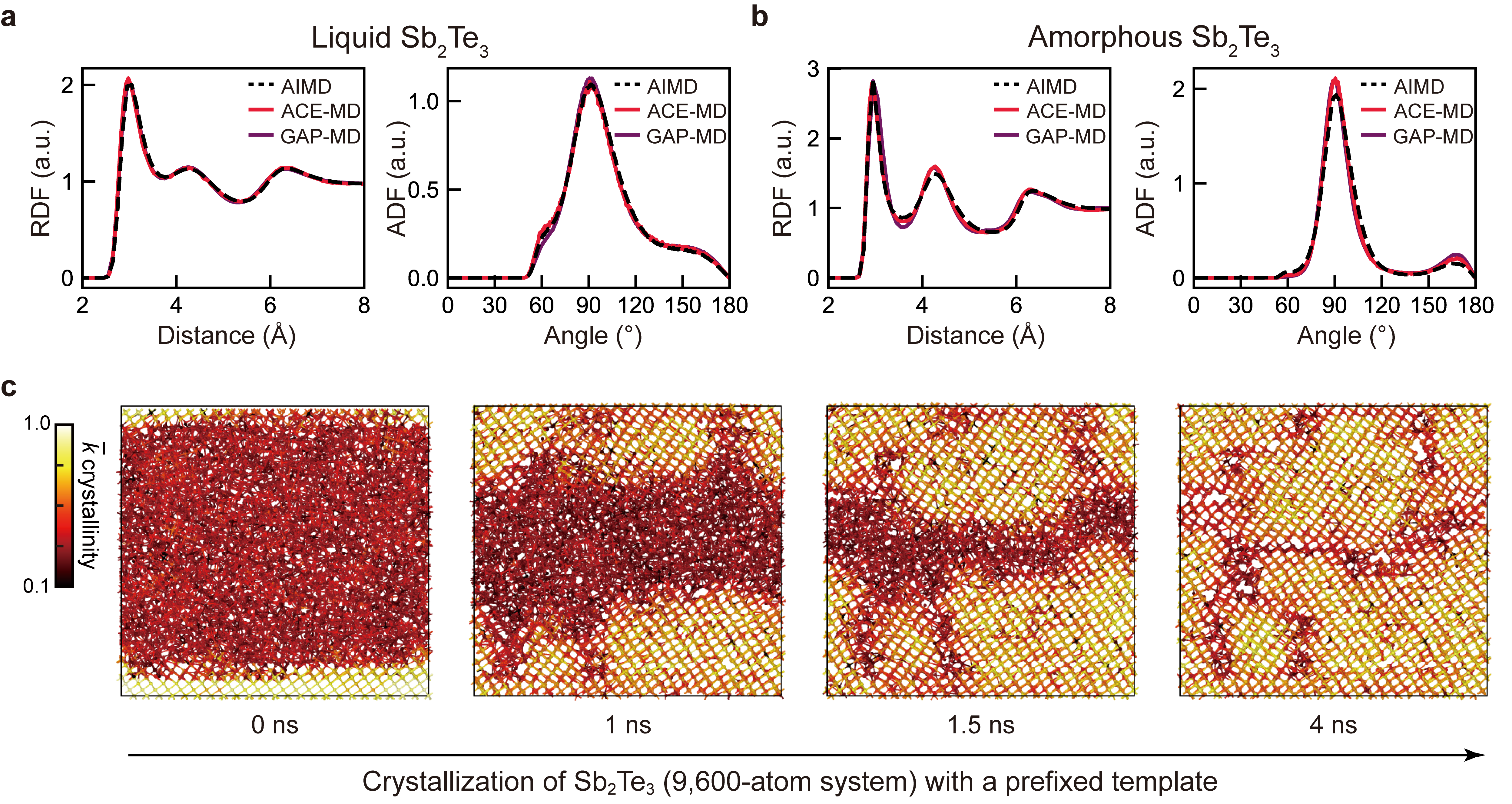}
   \caption{Structure and dynamics of \ce{Sb2Te3}. (\textbf{a}) RDF and ADF plots for liquid and amorphous structures simulated by the SbTe-XPOT-ACE (red) and GST-GAP-22 (purple) models as well as AIMD (black). The data for the latter two are taken from Ref.~\citenum{Zhou-23-10}. (\textbf{b}) Snapshots of an MD crystal-growth simulation. SOAP similarity is used to highlight the growth of crystalline \ce{Sb2Te3} (yellow) across snapshots. Several crystal grains and grain boundaries are visible.}
   \label{fig:sbte}
\end{figure*}

To test our approach for a more complex material system, we fitted a potential for $\rm{Sb_{2}Te_{3}}$, which is an important chalcogenide material used in various phase-change materials (PCM)-based devices for ultra-fast data storage \cite{Rao-17-12} and high-performance neuromorphic computing tasks. \cite{Ding-19-10, Zhang2019a} PCMs have long served as key application cases for ML potentials, including early work on the binary material GeTe \cite{Sosso2012} and the ternary \ce{Ge2Sb2Te5}. \cite{Mocanu2018} The former potential has been used for studies of crystallization \cite{Sosso2013} and thermal properties of GeTe; \cite{Sosso2012a} the latter has been applied to study structure and bonding in \ce{Ge2Sb2Te5} \cite{Konstantinou2019} and tested for the binary \ce{Sb2Te3}. \cite{Konstantinou2020}

Our previous work has introduced an ML potential based on the GAP framework for Ge--Sb--Te (GST) alloys located along the compositional tie-line between GeTe and \ce{Sb2Te3}. \cite{Zhou-23-10} This GAP model, which we call ``GST-GAP-22'', can accurately describe disordered structures of GST alloys and complex phase transition processes under practical programming conditions (e.g., non-isothermal heating) on the length scale of real-world devices. \cite{Zhou-23-10} We took a subset of the GST-GAP-22 dataset, which only contains elemental crystal structures of Sb and Te as well as binary bulk structures (including crystalline, amorphous, and intermediate crystallization configurations) found in the Sb--Te system.

A validation set was created using the protocol described in the Methods section, providing a representative sample of structures, and allowing us to consistently quantify the performance and robustness of our potentials. This is important for defining the loss function in XPOT (cf.\ Eq.~\ref{eq:loss}), and so the validation set is fixed for all potentials fitted during optimization.

Additionally, to confirm that the reduction in scope of the training set (from the full Ge--Sb--Te system to only the Sb--Te system) was not unfairly advantaging our own optimized potentials, we fitted a GAP with the same hyperparameters as for the GST-GAP-22 potential, \cite{Zhou-23-10} but using only the Sb--Te subset of the data for training. We use this potential (``SbTe-GAP'' in the following) as a benchmark as it provides very similar force errors to GST-GAP-22, but resulted in improved accuracy for energy prediction.

In this case, we cannot directly compare to GST-GAP-22 numerically, as the validation data are taken from the GST-GAP-22 training dataset. Therefore, to quantify the accuracy of the potentials, we created a new benchmark dataset similar to that built for silicon in Ref.~\citenum{George2020} (cf.\ Figure \ref{fig:si_gapmd_benchmark}). Specifically, we performed a GAP-MD simulation in which $\rm{Sb_{2}Te_{3}}$ was melted starting from a defective rocksalt-like crystal with an anion (Te) vacancy, before being quenched at $10^{13}~\mathrm{K~s}^{-1}$ to form the amorphous phase. This trajectory was then labeled with DFT to produce a benchmark set representing crystalline, liquid, and amorphous $\rm{Sb_{2}Te_{3}}$. Snapshots of these three phases are shown in Figure \ref{fig:sbte_gapmd}a. The atoms are color-coded by crystallinity as defined by the Smooth Overlap of Atomic Positions (SOAP)-based similarity, \cite{Bartok2013} with respect to rocksalt-like $\rm{Sb_{2}Te_{3}}$, as discussed in Ref.~\citenum{Xu-22-1}.

In Figure \ref{fig:sbte_gapmd}, we show that our XPOT-optimized ACE potential has improved not only upon the GST-GAP-22 predictions, but also the SbTe-GAP fitted on the Sb--Te subset of the full GST-GAP-22 database. The numerical accuracy is improved, and predictions are over $400 \times$ faster than for the GAP potentials. Figure \ref{fig:sbte_gapmd}b shows that the energy errors on the defective crystalline structures are consistently higher than for the liquid or amorphous phases across all potentials, with the same structural snapshots showing higher errors in this region. This occurs due to the relatively small number of crystalline structures with vacancy defects in the training database, and while our potentials all predict the energy to within 15 meV at.$^{-1}$, we see that these structures present more of a challenge to these ML potentials than the liquid or amorphous structures do.
The SbTe-XPOT-ACE potential is significantly more accurate than both GAPs in predicting energies and forces for amorphous structures, and the force predictions are uniformly over 10\% closer to DFT than those of by either GAP potential.

We carried out further physically-guided validation, testing the similarity of structural predictions between AIMD, GST-GAP-22, and our XPOT-ACE potential, aiming to verify that numerical errors correspond to well-described physical properties and processes in simulation. We first characterize the structure of amorphous $\rm{Sb_{2}Te_{3}}$. We ran a melt-quench simulation on a crystalline \ce{Sb2Te3} model containing 360 atoms. In Figure \ref{fig:sbte}a--b, we show that our liquid and amorphous structures generated using SbTe-XPOT-ACE closely match the RDF and ADF of structures generated using AIMD and GST-GAP-22, including the shoulder in the ADF of the liquid phase. Both GST-GAP-22 and the XPOT-ACE potential marginally over-order the amorphous phase of \ce{Sb2Te3}, evidenced by a slightly larger peak at 90° in the ADF and a larger second peak of the RDF, as compared to the AIMD reference (Figure \ref{fig:sbte}b).

To move beyond structural validation, we produced a 9,600-atom structural model of amorphous \ce{Sb2Te3} (in a box of 4.3 × 9.0 × 8.5 $\rm{nm^{3}}$) with a pre-fixed crystalline template to simulate the crystal-growth process. Figure \ref{fig:sbte}c shows a crystallization simulation for this templated structural model. As in Figure \ref{fig:sbte_gapmd}, we used SOAP-based similarity \cite{Bartok2013} with respect to rocksalt-like \ce{Sb2Te3} to quantify the per-atom crystallinity during the crystallization process. \cite{Xu-22-1} Our structural model was annealed at 600 K for 4 ns, and the growth proceeded quickly at the rough crystalline--amorphous interface. Upon nanosecond crystallization, we found many defects (e.g., point defects and layer stacking faults) in the recrystallized model (cf.\ the dark red atoms in Figure 5c), indicating competing growth of different crystalline regions with different crystal orientations. We note that such local disorder is challenging to fully characterize due to the short timescales on which they happen (e.g., in the programming operations of real-world devices), and high-temperature annealing can help to eliminate the local defects, e.g., via vacancy ordering, \cite{Zhang-16-5} resulting in an energetically more favorable crystalline phase with fewer defects. \cite{Wang-21-1}

\section*{Conclusions}

Hyperparameter optimization can improve the accuracy and efficiency of ML potentials. Alongside dataset construction and fitting architectures, the choice of hyperparameters remains an important aspect of fitting performant ML models. Well-chosen hyperparameters can lead to more accurate, and more efficient, ML potentials for accelerated materials modeling. 

In the present work, we have described an extension of our openly available XPOT code to ACE model fitting via \texttt{PACEMaker}.\cite{Lysogorskiy-21-6} We have shown example applications for Si and $\rm{Sb_{2}Te_{3}}$, two systems with diverse chemistry, across a wide range of configurational space (including liquid, amorphous, and crystalline phases). We thoroughly validated these potentials using a number of numerical and physical tests, and we believe that those types of benchmarks can be useful for other systems as well (see also Ref.~\citenum{Morrow-23-03}). In particular, expanding on Ref.~\citenum{ThomasduToit-23-6}, we studied how suitable different types of test sets are to assess the quality of ML potentials across a range of structures---including ``external'' test sets that are different from the validation set used during optimization. 

As ML potentials become larger and more complex, selecting suitable hyperparameters for fitting has become increasingly important to maximize performance and accuracy for a given training dataset. Furthermore, our results in Table \ref{tab:si_test} suggest that series of potential fits with systematically varied hyperparameters could help to diagnose possible areas of failures for candidate ML potentials: specifically, the XPOT-ACE-3 model has low numerical errors on static MQ-MD snapshots but fails when used to drive MD itself, and this may be correlated with an extremely high prediction error on the RSS test set, which is generated distinctly from both the physically motivated MQ-MD and test datasets. In this way, series of XPOT runs might help users to judge whether more data are required, or whether the model hyperparameters require improvement. We hope for this work to contribute to the wider uptake of ML potentials and their application to increasingly challenging research questions in chemistry and materials science.

\section*{Supporting Information}

Additional numerical results as a supplement to Table 1.

\section*{Acknowledgments}

We thank John Gardner, Minaam Qamar, Dr.~Anton Bochkarev, Dr.~Yury Lysogorskiy, and Prof.~Ralf Drautz for helpful discussions and Thomas Nicholas for testing early versions of the code.
This work was supported through a UK Research and Innovation Frontier Research grant [grant number EP/X016188/1].
Y.Z. acknowledges a China Scholarship Council-University of Oxford scholarship.
We are grateful for computational support from the UK national high performance computing service, ARCHER2, for which access was obtained via the UKCP consortium and funded by EPSRC grant ref EP/X035891/1.

\section*{Data and Code Availability}

The XPOT Python code is openly available via GitHub at
\href{https://github.com/dft-dutoit/xpot}{https://github.com/dft-dutoit/xpot}.
XPOT includes interfaces to third-party software, including \texttt{PACEmaker}\cite{Lysogorskiy-21-6} which is freely available for academic non-commercial use.
Data and code to reproduce the results from the present paper will be provided openly upon journal publication.

\providecommand{\latin}[1]{#1}
\makeatletter
\providecommand{\doi}
  {\begingroup\let\do\@makeother\dospecials
  \catcode`\{=1 \catcode`\}=2 \doi@aux}
\providecommand{\doi@aux}[1]{\endgroup\texttt{#1}}
\makeatother
\providecommand*\mcitethebibliography{\thebibliography}
\csname @ifundefined\endcsname{endmcitethebibliography}  {\let\endmcitethebibliography\endthebibliography}{}

\end{document}


\maketitle

In Table \ref{tab:init_late}, we show a comparison of the energy and force errors for each dataset. Upon optimization, we see an improvement in both energy and force prediction accuracy for the testing dataset across all values of $P$, excluding the XPOT-ACE-3 potential as documented in the main text.

In improving the energy accuracy of the testing set (through optimization), we observe increases in the energy error for the MQ-MD dataset, and inconsistent prediction accuracy on the RSS dataset. The prediction errors on the RSS dataset show that optimizing on the Si-GAP-18 test set does not necessarily improve the accuracy across all structures, and that in some cases the energy accuracy of predictions for the RSS dataset worsens despite simultaneously improving in accuracy on the MQ-MD dataset. However, for $P \le 2$, the RSS accuracy is always improved upon further optimization. The results for the ``4F" potentials further emphasize that there is not a direct link between force and energy error improvements.

\begin{table*}
 \caption{\label{tab:init_late} Energy and force RMSE values of silicon potentials either upfitted from the best initial potential (within the first 4 iterations, labelled Initial) or upfitted from the final optimised hyperparameters (labelled XPOT-ACE). The errors are evaluated on three different test sets. The number for the models refers to the number of atomic properties, $P$, from linear (1) to quaternary (4), excluding XPOT-ACE-6827, labeled after the number of functions in the potential. As in the main text, the test sets used are the Si-GAP-18 test set of Ref.~\citenum{Bartok2018} (``GAP-18'' for brevity), the melt--quench (MQ) MD test set of Ref.~\citenum{George2020}, and the random-structure-search (RSS) test set of Ref.~\citenum{Morrow-22-9}, respectively.}
 \fontsize{10.5pt}{14pt}\selectfont
 \begin{tabular}{lcc|ccc|ccc}
 \hline
 \hline
  & & & \multicolumn{3}{c}{Energy RMSE} & \multicolumn{3}{c}{Force RMSE} \\
  & & & \multicolumn{3}{c}{(meV at.$^{-1}$)} & \multicolumn{3}{c}{(meV \AA$^{-1}$)} \\
  \cline{2-5} \cline{5-9}
   & $P$ & \# Func. & GAP-18\cite{Bartok2018} & MQ-MD\cite{George2020} & RSS\cite{Morrow-22-9} & GAP-18 & MQ-MD & RSS \\
 \hline
 \multicolumn{9}{l}{\textit{XPOT-optimized models with increasing complexity}} \\
 \hline
 Initial-1 & 1 & 3000 & 4.2 & 4.0 & 28.7 & 76 & 110 & 160 \\
 XPOT-ACE-1 & 1 & 3000 & 3.5 & 5.1 & 27.1 & 69 & 105 & 158 \\
 \hline
 Initial-2 & 2 & 3000 & 6.5 & 3.5 & 192 & 70 & 104 & 1171 \\
 XPOT-ACE-2 & 2 & 3000 & 2.5 & 5.0 & 23.1 & 63 & 97 & 150 \\
 \hline
 Initial-3 & 3 & 3000 & 4.8 & 3.3 & 42.8 & 65 & 97 & 296 \\ 
 XPOT-ACE-3 & 3 & 3000 & 318 & 5.2 & $> 10^6$ & 300 & 98 & $> 10^8$ \\ 
 \hline
 Initial-4 & 4 & 3000 & 31.9 & 3.6 & 55.5 & 66 & 97 & 345 \\ 
 XPOT-ACE-4 & 4 & 3000 & 4.8 & 5.5 & 62.6 & 63 & 99 & 274 \\ 
 \hline
 \multicolumn{9}{l}{\textit{XPOT-optimized models with varied numbers of functions}} \\
 \hline
 Initial-3F & 3 & 1375 & 8.0 & 3.8 & 50.9 & 71 & 104 & 224 \\
 XPOT-ACE-3F & 3 & 2000 & 4.6 & 4.1 & 20.5 & 65 & 97 & 139 \\
 \hline
 Initial-4F & 4 & 875 & 3.7 & 5.4 & 47.0 & 71 & 103 & 240 \\
 XPOT-ACE-4F & 4 & 1625 & 2.5 & 5.4 & 72.5 & 64 & 100 & 187 \\
 \hline
 \multicolumn{9}{l}{\textit{Reference values}} \\
 \hline
 XPOT-ACE-6827 & 1 & 6827 & 3.0 & 4.4 & 34.5 & 63 & 104 & 179 \\
 REF-ACE (Ref.~\citenum{Lysogorskiy-21-6}) & 1 & 6827 & 3.2 & 4.3 & 42.1 & 77 & 124 & 175 \\
 \hline
 Si-GAP-18 (Ref.~\citenum{Bartok2018}) & --- & --- & 1.6 & 8.5 & 34.9 & 83 & 139 & 177 \\
 \hline 
 \hline
 \end{tabular}
\end{table*}

\providecommand{\latin}[1]{#1}
\makeatletter
\providecommand{\doi}
  {\begingroup\let\do\@makeother\dospecials
  \catcode`\{=1 \catcode`\}=2 \doi@aux}
\providecommand{\doi@aux}[1]{\endgroup\texttt{#1}}
\makeatother
\providecommand*\mcitethebibliography{\thebibliography}
\csname @ifundefined\endcsname{endmcitethebibliography}  {\let\endmcitethebibliography\endthebibliography}{}